# Characterization of NbTi wires for the electron-ion collider project

**Jun Lu[1*], Jeremy Levitan[1] Noah Gavin[1], Aniket Ingrole[1], Holger Witte[2], Peng Xu[2] and Ye Bai[2]**

[1] Magnetic Science and Technology, National High Magnetic Field Laboratory, Tallahassee, Florida, USA
[2] Brookhaven National Laboratory, Upton, New York, USA

[*]E-mail: junlu@magnet.fsu.edu

**Abstract.** The Electron-Ion Collider (EIC) is a proposed machine to explore the behaviour of the fundamental particles and forces that bind atomic nuclei together. The design and construction of the EIC are underway at Brookhaven National Laboratory in collaboration with Thomas Jefferson National Accelerator Facility. EIC will use several different types of superconducting strands for magnets near the interaction region (IR). At beam injection, the magnetic field is usually very low compared with its maximum operating field. This usually creates considerable field errors mainly generated from magnetization current in superconducting strands even using very fine filament. The accurate magnetization measurement results from those superconducting strands will be critical for the calculation and future correction of magnetic field for EIC.

In this work, we characterized three billets of superconductor NbTi strands. The magnetization was measured at 4.2 K and 1.9 K in magnetic fields below 1.5 T. The critical current at 4.2 K and in magnetic field down to 5 T were also measured. Other properties that are important for the safety margin of superconducting magnet fabrication, operation, and quench protection such as residual-resistance-ratio (RRR), filament diameter, Cu to non-Cu ratio, twist pitch, and mechanical properties at 77 K will also be presented.

## 1. Introduction

The Electron-Ion Collider (EIC) is a proposed machine to explore the behaviour of the fundamental particles and forces that bind atomic nuclei together [1]. In EIC, high energy and highly polarized hadron and electron beams will collide with a centre of mass energy up to 140 GeV to study the internal structure of proton and atomic nuclei [2]. The design and construction of the EIC are underway at Brookhaven National Laboratory (BNL) in collaboration with Thomas Jefferson National Accelerator Facility. EIC will use several different types of NbTi superconducting strands for large aperture dipole and quadrupole magnets near the interaction region (IR) of EIC. At beam injection, the magnetic field is usually very low compared with its maximum operating field. This usually creates considerable field errors mainly generated from persistent current in NbTi strands even using very fine filament. The accurate magnetization measurement results from those NbTi strands will be critical for the calculation and future correction of the magnetic fields. Other properties such as critical current, residual-resistance-ratio (RRR), filament diameter, twist pitch length and Cu/non-Cu ratio are also important parameter which impact the magnet design. In addition, the mechanical properties of NbTi wire at cryogenic temperatures are needed for mechanical design of magnet systems.

NbTi, a low Tc superconductor, has been widely used in accelerator magnets, fusion magnets, and magnets for magnetic resonance imaging (MRI). The test method for critical current [3] and other properties is well established [4]. In this paper, we present the results from the characterization tests of NbTi wire procured for EIC. It is also imperative to note that materials data such as these are valuable as part of the database to serve the community of superconducting magnet designers.

## 2. Experimental methods

Wires are fabricated by Bruker EAS. The wire diameter is 1.065 mm, with a nominal Cu/non-Cu ratio of 1.6, similar to type 01 strand used in the large hadron collider (LHC) at CERN [5]. For each of the test described below, except for the mechanical test, 3 samples were tested from each of the 3 billets.

Magnetization measurements are performed by a vibrating sample magnetometer (VSM) of the physical property measurement system (PPMS) made by Quantum Design Inc. Small 7-turn coils measuring approximately 6 mm in outer diameter are measured.

Critical Current ($I_c$) measurement is performed in a 15 T superconducting magnet. Sample wire is wound on a titanium test mandrel [6]. The sample is connected to a test probe using pressure contacts. The maximum allowed current of the probe is 1000 A. The criterion of 0.1 µV/cm is used to determine the $I_c$ with voltage tap length of 50 cm. During the measurement, the sample is immersed in liquid helium bath. No self-field corrections are made to the measured $I_c$ data.

Residual Resistivity Ratio (RRR) of NbTi wire is defined as ratio of resistance between 295 K and 10 K. Samples of 150 mm long are connected in series and placed on a G10 plate. A Cernox® temperature sensor is attached to the G10 plate using GE varnish. Resistances are measured by the four-probe method with current of +/- 1 A. 10 K temperature is reached by naturally warming up the system from 4.2 K. The typical warm up rate is about 10 mK per second. Further details of above measurement methods can be found in [6], [7].

The diameter of NbTi filaments is measured from optical micrograph taken from polished wire cross-sections. Micrographs are analysed using ImageJ software. The twist pitch is measured by the incline angle of the filaments, visible after etching away the outer copper stabilizer, with respect to the wire axial direction [6]. The Cu/non-Cu ratio is measured and calculated by copper weight loss after etching by nitric acid solution ($HNO_3$: $H_2O$ = 1:1) for several hours.

Mechanical properties are tested in liquid nitrogen using a universal test machine (MTS corporation). Aluminium grips designed for testing wires are used as shown in figure 1. A 25 mm clip-on extensometer (Epsilon Technology) calibrated at 77 K is used for strain measurements.

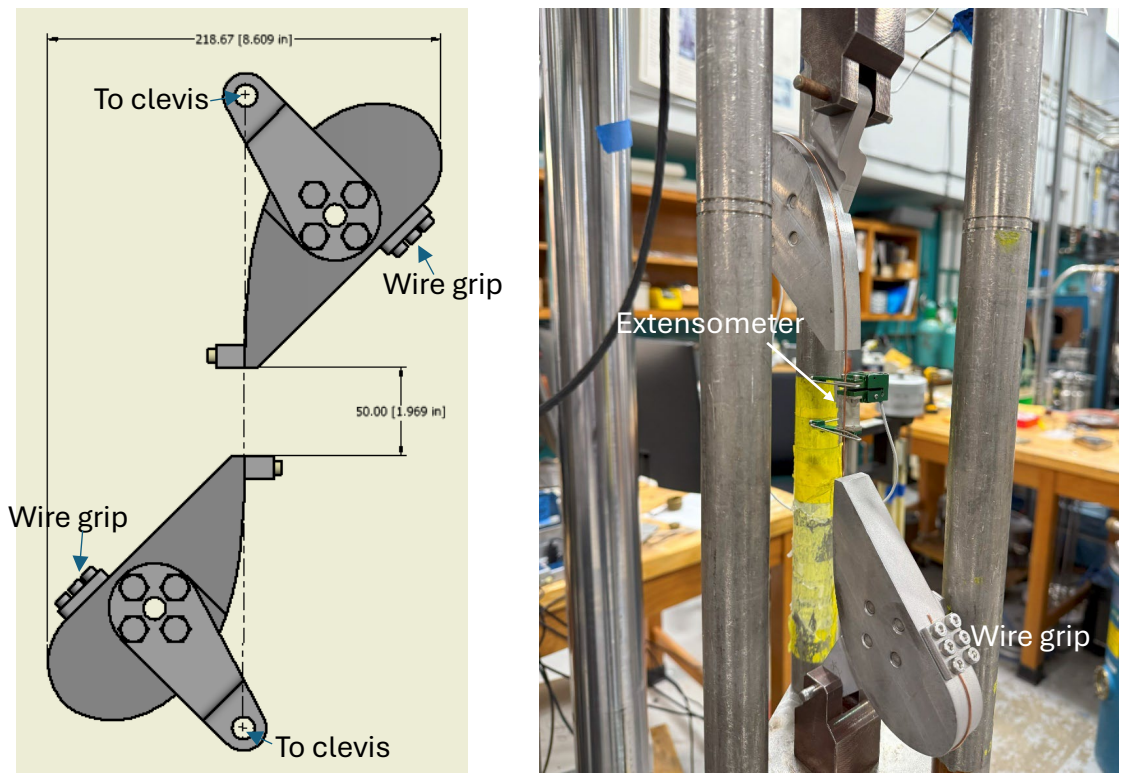

**Figure 1.** Mechanical testing of NbTi wire. (a) wire grip design. (b) a sample mounted on the grips with an extensometer on the MTS machine.

## 3. Results and discussions

### 3.1 Room temperature measurements

Figure 2 shows cross-section micrographs of two types of wires. The individual filaments are clearly resolved in Figure 2(c) and 2(d). After etching away the outer copper stabilizer, the twisting of filaments are revealed as shown in Figure 3. From the observed incline angle $\alpha$, the twist pitch is calculated. Table 1 summarizes the room temperature measurement results.

**Table 1.** Geometric characterization of NbTi wire.

| Sample ID | Filament diameter (μm) | | Cu/non-Cu ratio | | Twist pitch (mm) | |
|---|---|---|---|---|---|---|
| | Mean | stdev* | Mean | stdev* | Mean | stdev* |
| 113776AA | 9.8 | 1.5 | 1.59 | 0.02 | 18.7 | 1.2 |
| 114010AA | 6.5 | 1.4 | 1.67 | 0.01 | 19.7 | 0.6 |
| 114052BA | 6.5 | 1.7 | 1.69 | 0.01 | 21.4 | 1.6 |

*stdev = Standard Deviation

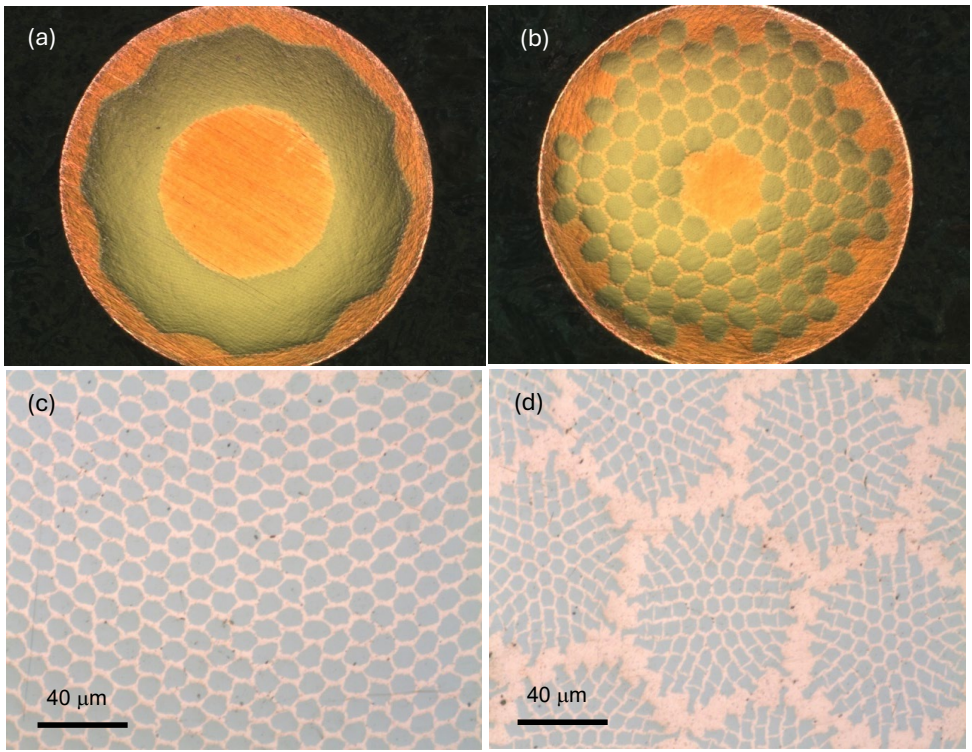

**Figure 2.** (a) and (b) cross-sections of 113776AA and 114010AA respectively. The architecture of 114052BA is the same as 114010AA. (c) and (d) the corresponding high magnification micrographs used to measure filament diameters.

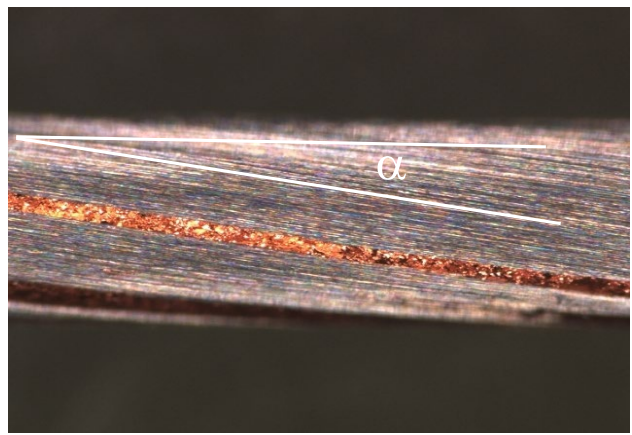

**Figure 3.** 113776AA wire after outer copper stabilizer is removed by nitric acid. The incline angle α is used to calculate twist pitch.

*3.2 Critical current and RRR*

The $I_c$ and RRR results are shown in Table 2. The RRR values are as expected for LHC strands of > 150 [8]. The variation in $I_c$ among the 3 tests for each billet is less than 1%. The average $I_c$ is plotted against the applied magnetic field as shown in figure 4. The measured $I_c$ values are slightly greater than LHC wire specification of > 532 A at 7 T and 4.2 K [9]. The specificatin of LHC wire $I_c$ is also ploted in figure 4 for comparison.

*3.3. Magnetization*

Figure 5 (a) is the initial magnetization curves up to 0.5 T for two different wires. Higher magnetization corresponds to the wire with larger filament diameter. For instance, the magnetization at 0.2 T for 113776AA is about 1.3 times larger than that of 114010AA, while the filament of 113776AA is about 1.5 times larger than that of 114010AA (Table 1). Figure 5 (b) compares the 0 – 1.5 T – 0 loops of 114010AA measured at 4.2 K and 1.9 K. The 2 K data from LHC type 01 NbTi strand is reproduced from [10] and plotted as the dash line for reference.

*3.4. Stress strain curves*

Figure 6 shows three stress strain curves from 114010AA samples at 77 K. An initial load of 50 N is applied to eliminate the error introduced by wire bending. The stess-strain curves can be seen to have a slope change at about 250 MPa. This is attributed to the yielding of Cu in the Cu-NbTi composite wire. Between 250 and 800 MPa, the stress-strain is approximately linear. The apparent modulus can be obtained in this quasilinear region, which is about 33 GPa. The resultant apparent modulus can be roughly explained by the rule of mixture of Cu and NbTi in the composite wire, where modulus of NbTi filaments is about 60 - 70 GPa at room temperature [11] and Cu is yielded at these stress levels. The ultimate tensile strength are about 1.0 GPa. Table 3 summarizes these properties with standard deviations as errors. One of three samples was unloaded from 0.01 to 0.005 strain then reloaded. Promonent hysteresis during the unload-reload is observed. Each test was terminated when the specimen factured which occured outside the extensometer.

**Table 2.** Critical current and RRR.

| Sample | RRR | 7.5 T | | 7 T | | 6.5 T | | 6 T | | 5.5 T | | 5.0 T | |
|---|---|---|---|---|---|---|---|---|---|---|---|---|---|
| | | $I_c$ | n | $I_c$ | n | $I_c$ | n | $I_c$ | n | $I_c$ | n | $I_c$ | n |
| 113776AA-1 | 209 | 469 | 26 | 577 | 30 | 688 | 34 | 796 | 38 | 907 | 38 | >1000 | |
| -2 | 214 | 469 | 28 | 577 | 32 | 688 | 36 | 799 | 39 | 907 | 44 | >1000 | |
| -3 | 208 | 469 | 25 | 580 | 31 | 688 | 34 | 799 | 39 | 907 | 38 | >1000 | |
| 114010AA-1 | 280 | 466 | 25 | 563 | 31 | 664 | 29 | 765 | 35 | 865 | 30 | 958 | 30 |
| -2 | 276 | 475 | 26 | 568 | 27 | 673 | 30 | 766 | 31 | 868 | 32 | 961 | 30 |
| -3 | 279 | 466 | 29 | 565 | 29 | 664 | 33 | 766 | 34 | 865 | 35 | 961 | 35 |
| 114052BA-1 | 209 | 448 | 24 | 550 | 28 | 646 | 29 | 742 | 31 | 838 | 32 | 934 | 30 |
| -2 | 201 | 451 | 26 | 550 | 29 | 649 | 30 | 745 | 32 | 841 | 33 | 929 | Q |
| -3 | 204 | 454 | 25 | 553 | 27 | 649 | 29 | 745 | 30 | 841 | 32 | 929 | 31 |

*Q is quenched before transition.

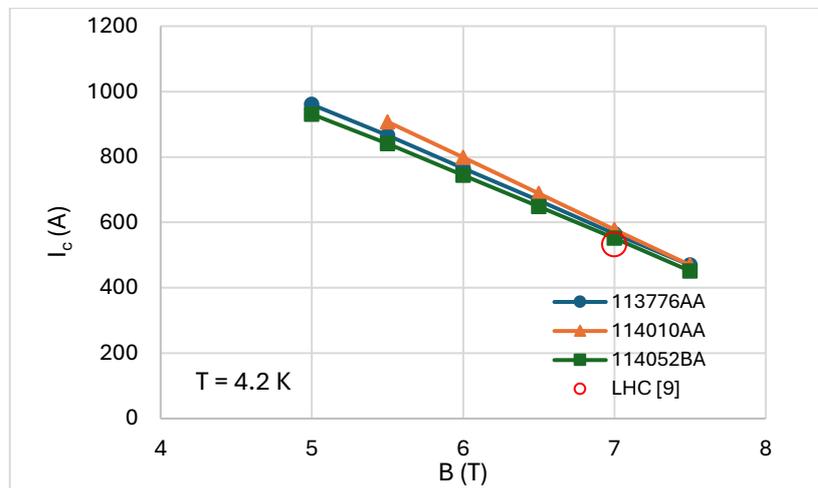

**Figure 4.** Average $I_c$ versus B of samples from 3 billets. Specification of LHC type 01 wire minimum $I_c$ [9] is also plotted as the empty circle for reference.

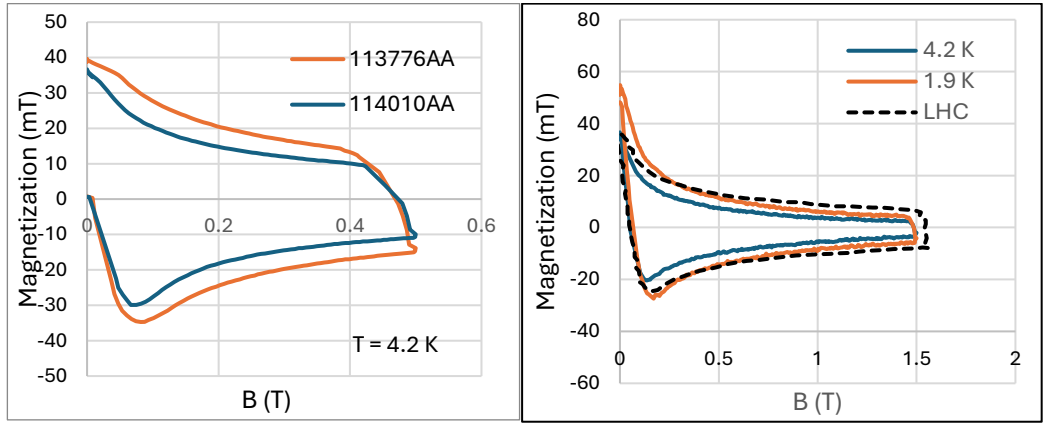

**Figure 5.** Magnetization versus applied magnetic field (a) the initial magnetization (0 – 0.5 -0 T) comparison of wires with different design. Larger filament diameter corresponds to higher magnetization. (b) 1.5 - 0 -1.5 T at 4.2 K and 1.9 K of 114010AA wire. The magnetization of LHC type 01 wire at 2 K is reproduced from [10] and plotted for comparison.

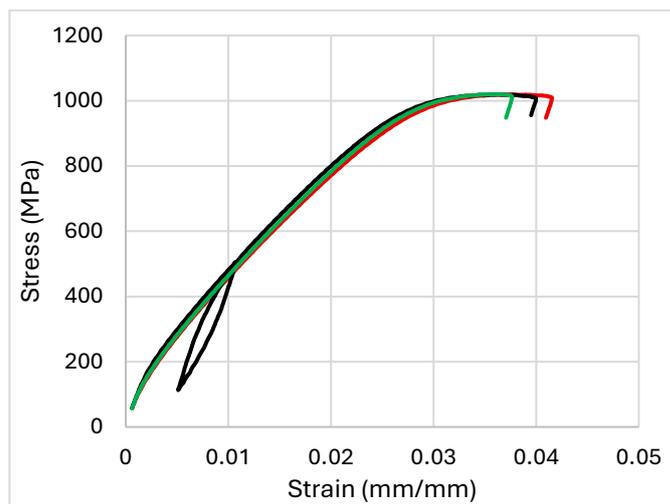

**Figure 6.** Stress-strain curves of 1114010AA NbTi wires (3) at 77 K.

**Table 3.** Mechanical properties of NbTi wire at 77 K

| ID | # of test | Apparent Modulus* (GPa) | UTS (MPa) | Strain at failure (%) |
|---|---|---|---|---|
| 113776AA | 1 | 33.5 | 972 | 3.9 |
| 114010AA | 3 | 32.6 ± 0.9 | 1020 ± 0.4 | 3.9 ± 0.2 |
| 114052BA | 2 | 33.3 ± 0.1 | 1017 ± 5.5 | 3.8 ± 0.01 |

*Apparent modulus is obtained from data between 250 and 800 MPa. The errors presented are standard deviations.

## 4. Conclusions

In this work, magnetization, critical current, RRR, filament diameter, Cu/non-Cu ratio, twist pitch and the mechanical properties of NbTi wires used for EIC design studies are characterized. Magnetization, $I_c$ at 4.2 K, RRR, filament diameter, twist pitch are comparable to the type 01 wires of the LHC. The magnetization is measured at both 4.2 K and 1.9 K. Larger filament diameter corresponds to higher magnetization. Mechanical testing is performed at 77 K. The ultimate tensile strength is about 1.0 GPa. The apparent modulus between 250 and 800 MPa is 33 GPa.

## Acknowledgments

We thank Dr. Ke Han for helpful discussions on mechanical data analysis, Mike White for modification of the wire grips. This work was supported by Brookhaven Science Associates, LLC under contract No. DE-SC0012704 with the U.S. Department of Energy. This work was performed at the National High Magnetic Field Laboratory, which is supported by National Science Foundation Cooperative Agreement No. DMR- 2128556, and the State of Florida.

## References

[1] F. Willeke, Electron ion collider conceptual design report, Brookhaven National Laboratory, 2021.
[2] A. Deshpande, Joint 20th International Workshop on Hadron Structure and Spectroscopy and 5th workshop on Correlations in Partonic and Hadronic Interactions, Sept. 2024, Yerevan, Armenia.
[3] L. F. Goodrich, et al., NBS special publication 260-91, 1984.
[4] I. Pong, et al., IEEE Trans. Appl. Supercond., **22** (3), 4802606, (2012).
[5] J. D. Adam, et al., IEEE Trans. Appl. Supercond., **12** (1), 1056, (2002).
[6] D. McGuire, et al., IEEE Trans. Appl. Supercond., **25** (3), 9500304, (2015).
[7] J. W. Levitan, et al., IEEE Trans. Appl. Supercond., **29** (5), 6000904, (2019).
[8] Z. Charifoulline, IEEE Trans. Appl. Supercond., **16** (2), 1188, (2006).
[9] J. Fleiter, TE/MSC Seminar, CERN, Sept 2021
[10] S. Le Naour, et al., IEEE Trans. Appl. Supercond., **9** (2), 1763, (1999).
[11] Tomomichi Ozaki, et al., Materials Transactions, **45** (8), 2776, (2004).